\documentclass[sn-standardnature]{sn-jnl}

\usepackage{appendix} 
\usepackage{caption}

\usepackage{amsmath}
\usepackage{xcolor}
\usepackage{braket} 
\usepackage{float}
\usepackage{graphicx}
\usepackage{times}

\usepackage{physics}
\usepackage{geometry}

\theoremstyle{thmstyleone}%
\theoremstyle{thmstyletwo}%

\theoremstyle{thmstylethree}%

\begin{document}

\title[Germanene Nanoribbons]{Realization of a one-dimensional topological insulator in ultrathin germanene nanoribbons}

\author[1]{\fnm{Dennis J.} \sur{Klaassen}}\equalcont{These authors contributed equally to this work.}

\author[2]{\fnm{Lumen A.G.} \sur{Eek}}\equalcont{These authors contributed equally to this work.}

\author[3]{\fnm{Alexander N.} \sur{Rudenko}}

\author[1]{\fnm{Esra D.} \sur{van't Westende}}

\author[1]{\fnm{Carolien} \sur{Castenmiller}}

\author[1,4]{\fnm{Zhiguo} \sur{Zhang}}

\author[1]{\fnm{Paul} \sur{de Boeij}}

\author[1]{\fnm{Arie} \sur{van Houselt}}

\author[5]{\fnm{Motohiko} \sur{Ezawa}}

\author[1]{\fnm{Harold J.W.} \sur{Zandvliet}}

\author*[2]{\fnm{Cristiane} \sur{Morais Smith}}\email{c.demoraissmith@uu.nl}

\author*[1]{\fnm{Pantelis} \sur{Bampoulis}}\email{p.bampoulis@utwente.nl}

\affil[1]{\orgdiv{Physics of Interfaces and Nanomaterials, MESA+ Institute for Nanotechnology}, \orgname{University of Twente}, \orgaddress{\street{Drienerlolaan 5}, \city{Enschede}, \postcode{7522 NB}, \country{the Netherlands}}}

\affil[2]{\orgdiv{Institute for Theoretical Physics}, \orgname{Utrecht University}, \orgaddress{\street{Princetonplein 5}, \city{Utrecht}, \postcode{3584 CC}, \country{the Netherlands}}}

\affil[3]{\orgdiv{Institute for Molecules and Materials}, \orgname{Radboud University Nijmegen}, \orgaddress{\street{Heyendaalseweg 135}, \city{Nijmegen}, \postcode{6525 AJ}, \country{the Netherlands}}}

\affil[4]{\orgdiv{Department of Materials Science}, \orgname{Fudan University}, \orgaddress{\street{220 Handan Road}, \city{Shanghai}, \postcode{200433}, \country{China}}}

\affil[5]{\orgdiv{Department of Applied Physics}, \orgname{University of Tokyo}, \orgaddress{\street{Hongo}, \city{Tokyo}, \postcode{113-8656}, \country{Japan}}}

\abstract{
Realizing a one-dimensional (1D) topological insulator and identifying the lower-dimensional limit of two-dimensional (2D) behavior are crucial steps toward developing high-density quantum state networks, advancing topological quantum computing, and exploring dimensionality effects in topological materials. Although 2D topological insulators have been experimentally realized, their lower dimensional limit and 1D counterparts remain elusive. Here, we fabricated and characterized arrays of zigzag-terminated germanene nanoribbons, a 2D topological insulator with a large topological bulk gap. The electronic properties of these nanoribbons strongly depend on their width, with topological edge states persisting down to a critical width ($\sim$ 2 nm), defining the limit of 2D topology. Below this threshold, contrary to the tenfold way classification, we observe zero-dimensional (0D) states localized at the ends of the ultrathin nanoribbons. These end states, topologically protected by time-reversal and mirror symmetries, mark the first realization of a 1D topological insulator with strong spin-orbit coupling. Our findings establish germanene nanoribbons as a platform for investigating 1D topology and dimensionality effects in topological materials.
}


\maketitle

\clearpage
\newpage
Low-dimensional topological insulators (TIs) have emerged as prime candidates for next-generation quantum technologies, particularly in low-energy electronics and quantum computing. This is because of their unique electronic properties: an insulating bulk paired with symmetry-protected boundary states \cite{kane2005quantum, kane2005z, bernevig2006quantum, bernevig2006quantumB, hasan2010colloquium, wu2006helical, qi2011topological, kempkes2019robust}. Significant progress has been made in discovering and characterizing various two-dimensional (2D) TIs hosting one-dimensional (1D) dissipationless edge states, such as band-inverted semiconductors, bismuthene, WTe$_2$, and germanene \cite{konig2007quantum, roth2009nonlocal, qian2014quantum, li2015observation, beugeling2015topological, knez2011evidence, wu2018observation, tang2017quantum, fei2017edge, reis2017bismuthene, collins2018electric, bampoulis2023quantum, lodge2021atomically}. To harness the potential of these 1D edge states in devices, it is essential to maximize their density per unit area \cite{gilbert2021topological, weber20242024}, which can be achieved by fabricating arrays of narrow nanoribbons. However, there may be a critical width below which the system transitions from a 2D to a 1D topological insulator, leading to the disappearance of the 1D edge states. This behavior is analogous to the dimensional crossovers observed when reducing three-dimensional (3D) TIs and semimetals to two dimensions \cite{zhang2010crossover, beugeling2012topological, collins2018electric, xia2019dimensional}.

While 2D and 3D TIs exhibit conductive channels at their edges or surfaces, 1D TIs are characterized by 0D states at their ends. Realizing a 1D TI is both a major challenge and a promising avenue for advancing topological quantum computing, where protected 0D end states could enable fault-tolerant qubit designs \cite{rizzo2018topological, alicea2012new, sarma2015majorana, weber20242024}. Although theoretical models \cite{su1979solitons, marques2019one, guo2016brief, jin20201d, pham2020emergence, liu2022ta} and quasi-1D structures, including graphene nanoribbons \cite{autes2016novel, traverso2024emerging, rizzo2018topological}, hint at feasibility, there are currently no experimentally verified 1D TIs with large spin-orbit coupling. Therefore, investigating the transition from 2D to 1D topology is a critical step, essential not only for forming dense networks of 1D topological states but also for realizing the 1D TI with symmetry-protected 0D end states.

In this study, we fabricated and characterized arrays of narrow, straight, and long germanene nanoribbons, a platform for achieving lateral arrangements of multiple 1D edge states in a system with large spin-orbit coupling. Using scanning tunneling microscopy (STM) and theoretical analysis, we explore how the topology of germanene nanoribbons varies with width in the framework of the tenfold way classification \cite{ryu2010topological, chiu2016classification}. We identified the lower-dimensional limit of 2D topology, laying the foundation for scalable topological quantum systems, and realized the long-sought 1D topological insulator with robust, symmetry-protected end states.

Germanene, a monolayer of germanium atoms in a low-buckled honeycomb lattice (Fig. 1\textbf{a}) \cite{cahangirov2009two, bampoulis2014germanene, bechstedt2021beyond, molle2017buckled, zhang2016structural}, has a spin-orbit coupling (SOC)-induced band gap at the K and K' points, Fig. 1\textbf{b}, and is classified as a Kane-Mele-type TI (class AII) with topologically protected helical edge states \cite{kane2005quantum, kane2005z}. The topological properties of 2D germanene were recently experimentally confirmed \cite{bampoulis2023quantum, zandvliet2024evidence}. To explore these properties in one-dimensional confinement, we fabricated germanene nanoribbons by depositing $\sim$1 monolayer of Pt on a Ge(110) substrate, then annealing at 1100 K. Upon cooling, Ge segregates on the Pt layer forming elongated nanoribbons, a growth process similar to the growth of germanene nanoribbons on Ag films on Ge(110) \cite{yuhara2021epitaxial}. For details, see the sample preparation section in Supplementary Information (SI). Two distinct structures emerge based on Pt coverage: short (10–50 nm), disordered Ge nanowires (small Pt coverages), and long (up to 1 $\mu$m), ordered nanoribbons (larger Pt coverages), oriented along the [$\bar{1}$10] direction. The nanoribbons have unique electronic properties and structural order, contrary to the disordered and electronically featureless nanowires (see SI). Fig. 1\textbf{c} shows a region with arrays of nanoribbons and nanowires, caused by local inhomogeneities in Pt coverage. An array of nanoribbons appears on the right, whereas a few isolated nanoribbons, surrounded by nanowires, are visible on the left of the STM image.

Consistent with 2D germanene, atomic resolution STM images on the nanoribbons (Figs. 1\textbf{d-f}) reveal the honeycomb germanene lattice with a lattice constant of $\sim$4.2 \AA~ \cite{bampoulis2014germanene, cahangirov2009two}. Micro-spot low-energy electron diffraction ($\mu$LEED) confirms this structure (see SI). A monolayer of germanium forms a low-buckled honeycomb structure (germanene) due to the instability of high-buckled Ge(111), which exhibits imaginary phonon modes in a large part of the Brillouin zone \cite{cahangirov2009two}. The honeycomb lattice, contrasting the rectangular Pt/Ge(110) substrate (similarly to silicene nanoribbons on Ag(110) \cite{de2010evidence, cerda2016unveiling}), indicates weak germanene-substrate interactions. The germanene nanoribbons align perfectly along the [$\bar{1}$10] direction but experience a lattice mismatch with the substrate along [001], resulting in very narrow widths (1 - 6 nm) (Fig. 1\textbf{d}) \cite{tersoff1993shape}. STM images (Figs. 1\textbf{e} and \textbf{f}) show that the nanoribbons terminate in zigzag edges along the [$\bar{1}$10] direction, consistent with the lower formation energy of zigzag over armchair edges. Moreover, we observe a buckling with an upper bound of $\sim$0.35 \AA~ (Fig. 1\textbf{g}). Fig. 1\textbf{h} shows the proposed structural model.

The formation of germanene nanoribbons opens up new avenues for investigating electronic behaviors distinct from those in extended 2D layers. To this end, we examined the spatial variation in the local density of states (LDOS) of a wide germanene nanoribbon ($\sim$6 nm) using scanning tunneling spectroscopy (STS). Fig. 2\textbf{a} shows the local differential conductivity ($dI(V)/dV$) spectra recorded at the edge (red curve) and the middle (black curve) of the nanoribbon shown in the inset. A distinct difference between edge and bulk is evident, with a pronounced edge state at $\sim$30 meV, centered within the bulk gap. Due to proximity effects, the underlying Pt containing layer increases the topological bandgap of germanene to 100–150 meV (width of the \textit{dI/dV} dip) at a germanene–Pt separation of 3.2 \AA~\cite{bampoulis2023quantum}, which exceeds that of free-standing germanene \cite{liu2011quantum}. Note that the gap does not always reach zero, probably due to the tip convolution effects from the edge states or the influence of the Pt layer lifting the $dI/dV$ baseline.

Fig. 2\textbf{b} shows a $dI(V)/dV$ line spectroscopy along the blue arrow in the inset of Fig. 2\textbf{a}, highlighting edge states on both sides of the nanoribbon (marked by red arrows). The metallic edge states are shifted relative to each other by about 20-30 meV, indicating a different local electrostatic environment. The intensity of these states also varies, possibly due to the buckling. $\textit{dI/dV}$ mapping of the edge state, shown in Fig. 2\textbf{c} for a wide ribbon, reveals that the state runs continuously along the edge of the nanoribbon. The topological nature of these edge states is further confirmed by tip-induced electric field measurements \cite{drummond2012electrically, ezawa2013quantized, bampoulis2023quantum}. By applying a perpendicular STM tip-induced electric field, we break band inversion symmetry, converting germanene into a trivial insulator devoid of edge states, as shown in Fig. 2\textbf{d}. Here, increasing the current set-point (thus reducing the tip-sample distance) increases the tip-induced electric field, which opens a gap in the energy of the edge states under strong fields. This gap closes and the edge states reappear when the electric field is reduced to its original value, confirming that the transition is reversible. The absence of such behavior in trivial surface states strongly reinforces the topological origin of these edge states, as does the reversible electric-field-induced transition, consistent with previous studies \cite{bampoulis2023quantum, collins2018electric}.

The reversible tuning of edge states with an applied electric field highlights the unique topological characteristics of germanene nanoribbons. This behavior invites a closer examination of how the edge states depend on the width of the nanoribbon. The natural width variations of the nanoribbons observed in STM images provide an ideal opportunity to explore their width-dependent electronic properties. Fig. 2\textbf{e} depicts $dI(V)/dV$ point spectra recorded at the interior (black) and edges (red) of nanoribbons with varying widths, namely $\sim$1 nm, $\sim$1.5 nm, $\sim$2.4 nm, $\sim$3.6 nm, and $\sim$6.0 nm. Strikingly, the presence of edge states strongly depends on the nanoribbon width. Nanoribbons wider than approximately 2.4 nm consistently display pronounced edge states, while narrower ribbons (below this threshold) exhibit nearly identical \textit{dI(V)/dV} spectra at both the interior and edges, with a marked absence of edge states.

To understand this, we performed tight-binding calculations on zigzag germanene nanoribbons using the Kane-Mele model \cite{kane2005quantum, kane2005z} for spin-orbit coupling strength $\lambda_{SO} = 0.3t$, staggered mass $M_{s} = 0.02t$, long-range hopping $t_{3} = 0.3t$, and nearest-neighbor hopping $t = 0.92$ eV \cite{ezawa2012topological, ezawa2013quantized}. For details on the theoretical calculations, see SI. The band structure is shown in Figs. 2\textbf{f} and 2\textbf{g}. A wide nanoribbon (14 hexagonal cells wide), showcases pronounced edge states and a bulk gap, see Fig. 2\textbf{f}. However, our band structure calculations of a 2-hexagon-wide nanoribbon reveal that the topological 1D edge modes fail to completely close the bulk gap, leading to gapped edge states, see Fig. 2\textbf{g}. This striking observation indicates a width-driven topological phase transition. Moreover, our theoretical results show the presence of 0D end states within this edge gap (or minigap), see Fig. 2\textbf{g}.

To examine this phase transition and the emergence of distinct end states in ultrathin nanoribbons, we conducted a detailed examination of the width-dependent electronic structure across various nanoribbon widths: wide ($\sim$5 nm, or about 14 hexagons), intermediate ($\sim$2.4 nm, or 6 hexagons), 3-hexagon-wide ($\sim$1.5 nm), and 2-hexagon-wide ($\sim$1 nm) nanoribbons. Our study encompasses both theoretical (Figs. 3\textbf{a}, \textbf{b}, \textbf{c}, and \textbf{d}) and experimental (Figs. 3\textbf{e}, \textbf{f}, \textbf{g}, and \textbf{h}) analyses. Spectra were recorded along various regions of the nanoribbons, where solid red lines indicate spectra from the top and bottom edges, black lines from the bulk, and blue lines from the ends (see inset of Figs. 3\textbf{e-h} for the corresponding topography). The theoretical LDOS curves are calculated for all energies in the bulk gap. As a result, bulk bands start to contribute around $E \approx \pm t$, yielding an increased LDOS. As we will discuss next, the theoretical description captures the main experimental features. For completeness, we also provide the corresponding \textit{dI(V)/dV} line spectroscopy along the nanoribbon end for all four nanoribbon widths in Figs. 3\textbf{i}, \textbf{j}, \textbf{k}, and \textbf{l}. 

Experimental \textit{dI(V)/dV} curves for the wide nanoribbon ($\sim$5 nm) reveal similar intensities of in-gap states at both edges and ends (Fig. 3\textbf{e}, blue and red curves), consistent with Fig. 2\textbf{a} and with the properties of a 2D TI. Note that the peaked LDOS lacks physical meaning. For a 1D edge state, the LDOS should be constant, as shown in Fig. 3\textbf{a}. This discrepancy arises from the influence of $k_{//}$ (parallel momentum) in STS. The zigzag edge state connects the K and K' points through the $\Gamma$ point. In STS, the largest tunneling contributions come from states at the $\Gamma$ point and decrease away from this point. As sample bias increases, we move away from the $\Gamma$ point and the signal decreases, leading to a peaked LDOS (for details see Ref. \cite{zandvliet2024evidence}). Spin-orbit interactions further modulate the dispersion, enhancing the effect \cite{matthes2014influence}. 

In contrast, the 6-hexagon-wide nanoribbon (Fig. 3\textbf{f}) has reduced LDOS for edge states (red) compared to the more prominent and sharper end state (blue). Strikingly, in the 3-hexagon-wide nanoribbon (Fig. 3\textbf{g}), both edge and end states vanish entirely. This observation is supported by our theoretical findings (Fig. 3\textbf{c}), which will be further elaborated upon later. Remarkably, the 2-hexagon wide nanoribbon (Fig. 3\textbf{h}), although also devoid of edge states, exhibits a very strong and pronounced end state (blue curve). This is unexpected, as class AII systems do not exhibit topological states in 1D. However, as demonstrated in the SI, these 0D end states are robust and topologically protected by the combined action of mirror and time-reversal symmetries. The emergence of the 0D end state in our experiments is fully supported by our theoretical calculations (Fig. 3\textbf{d}), where the blue curve (recorded at the end) displays a clear peak, signaling a 0D end mode, while the red and black curves show a dip. However, unlike the experimental curves, the calculated LDOS for 1D edge states does not completely vanish. The tight-binding calculations may underestimate the decay length of the 1D edge states, such that they fail to hybridize and disappear. The topography and \textit{dI/dV} maps of an ultrathin (2-hexagon wide) nanoribbon at energies near and far from the end state energy ($\sim$ 75 meV) are shown in Fig. 3\textbf{m}, highlighting the end state localization at the end of the nanoribbon. Finally, the difference between the 2- and 3-hexagon wide nanoribbons, i.e., an end mode for the former and no end-mode for the latter, is reminiscent of the different behavior predicted by Traverso \textit{et al.} \cite{traverso2024emerging} for graphene nanoribbons with even/odd number of hexagons. Their analysis was performed for a Haldane model, but since the Kane-Mele model consists of two time-reversal symmetric copies of the Haldane model, it can be applied also here. 

The experimental observations and the theoretical LDOS curves suggest a critical width threshold below which end states emerge. In germanene nanoribbons, this corresponds to six hexagons. When sufficiently thin, nanoribbons transition to a 1D topological insulator, hosting 0D end modes residing in the gapped edge states. Indeed, the presence of 0D topological end states in the 2- and 6-hexagon wide nanoribbons is confirmed by a finite topological invariant obtained from the 1D Zak phase \cite{zak1989berry}, see SI. The 0D topological states are two exponentially localized modes at the ends of the nanoribbons. The minigap's topological properties and size vary with nanoribbon width. The gap size scales inversely with width, and the end mode localization length scales inversely with gap size (see SI). Consequently, as we increase the ribbon width, the distinction between 0D end states and crossing 1D edge states becomes increasingly difficult. For wider nanoribbons, the 0D end states merge into the 1D edge states and are no longer visible. The 6-hexagon wide nanoribbon is an intermediate case, hosting both 1D edge states and 0D end states. In the SI, we show that nanoribbons of odd width (odd number of hexagons) always show a gap in the edge state spectrum, although this gap becomes exponentially small as we increase their width. This is in contrast with nanoribbons of even width, which have a gapless edge state spectrum. A staggered mass opens these gaps and reveals the 0D end states. Therefore, the topological behavior of the nanoribbons strongly depends on their width. This is exemplified in Figs. S7\textbf{d-e}, which show that the nanoribbons that are 2 hexagonal cells wide are topological for a larger parameter range than the nanoribbons that are 3 hexagonal cells wide. This is directly corroborated by the tight-binding and experimental results shown in Figs. 3\textbf{c} and \textbf{g}, where the nanoribbon is in its trivial state and has no end state. These findings emphasize the role of physical dimensions in the manifestation of topological properties. 

In conclusion, our theoretical and experimental investigation reveals the complex and rich topological behavior of germanene nanoribbons. We observe a transition from a class AII 2D quantum spin Hall phase to a 1D topological crystalline phase. We demonstrated that the 1D topological phase is influenced in a non-monotonic manner by many factors, i.e., ribbon width, spin-orbit coupling, next-nearest-neighbor hopping, and staggered mass. These findings are crucial for understanding the transition from a 2D to a 1D topological insulator as a function of width, highlighting the significance of physical dimensions on topological properties and quantitatively defining the maximum number of 1D edge modes per sample area in germanene nanoribbon arrays. Furthermore, our results show that sufficiently narrow nanoribbons with mirror-symmetric termination can support end states of fractionalized electrons \cite{benalcazar2019quantization, van2018higher, fang2019new, li2020fractional}, akin to Majorana zero modes, positioning them as a promising platform for exploring unique topological phases.

\section*{Data availability}
The experimental data that support the findings of this study are available from the corresponding authors upon reasonable request. 

\bibliography{Bibliography_EFTT} 

\subsection*{Acknowledgments}
 DJK, CC, HJWZ, and PB acknowledge NWO Veni and NWO Grants 16PR3237 and OCENW.M20.232 for financial support. LAGE, HJWZ, and CMS  acknowledge QUMAT for financial support. ME acknowledges CREST, JST (JPMJCR20T2), and Grants-in-Aid for Scientific Research from MEXT KAKENHI (Grant No. 23H00171) for financial support. ZZ thanks CSC for financial support. LAGE thanks Anouar Moustaj for the fruitful discussions on topology.

\subsection*{Author contributions}
DJK and LAGE contributed equally to this work. DJK, EDvW, and CC carried out the STM experiments. DJK and PB analyzed the STM data. PB supervised the STM experiments. LAGE performed the tight-binding calculations. CMS supervised the tight-binding calculations. PB and CMS wrote the paper. ZZ and AvH did the LEEM experiments. All authors contributed to the development of the experimental and theoretical concepts, interpretation of the data and reviewed the paper.

\subsection*{Competing interests} The authors declare no competing interests.

\subsection*{Additional information}
\textbf{Supplementary information.} The online version contains supplementary material available at...

\newpage

\begin{figure*}[tb]
	\centering
	\includegraphics[width=0.95\textwidth]{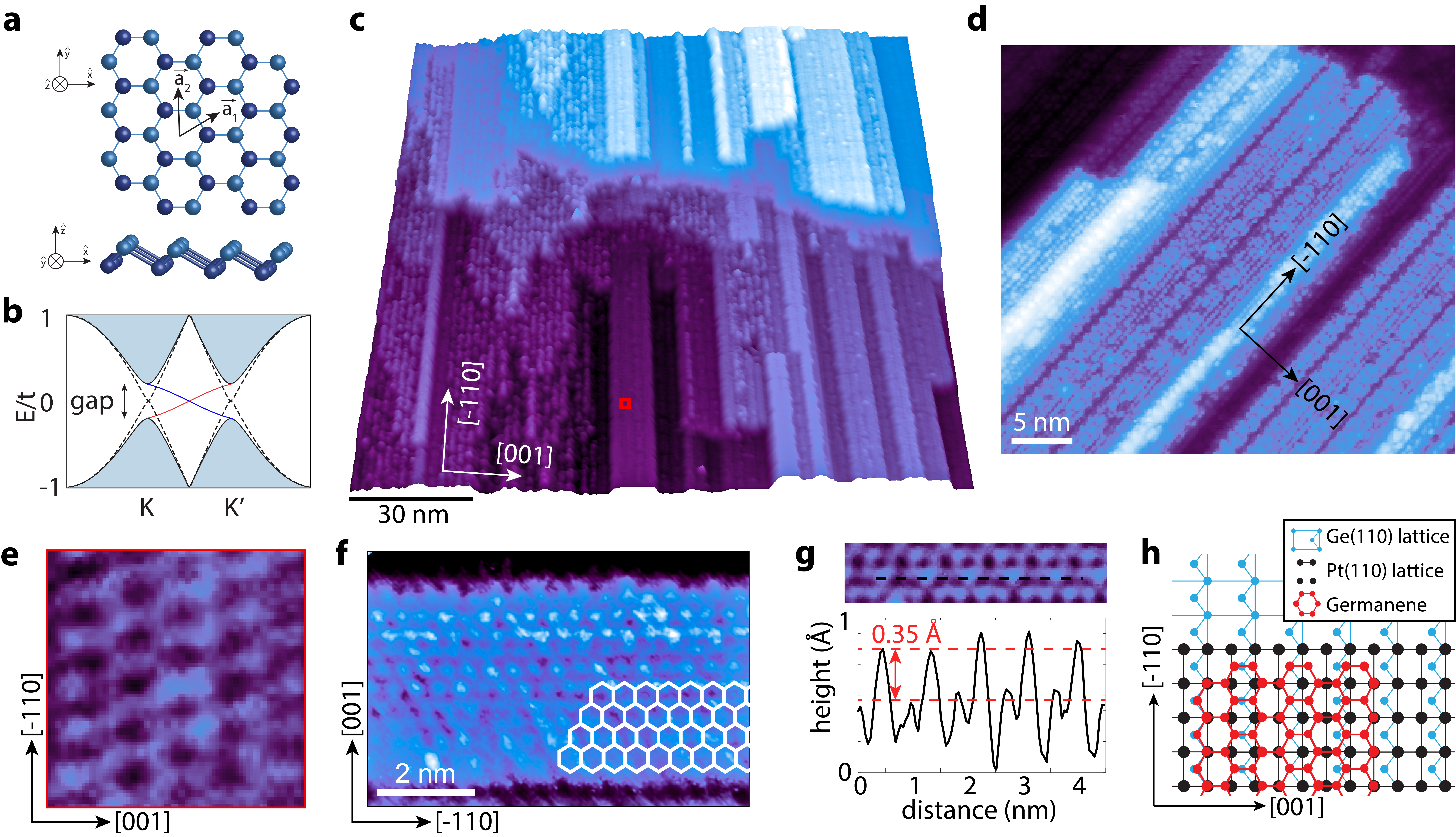}
	\caption{\textbf{Germanene nanoribbon structure.} 
\textbf{a} Top- and side-view of germanene’s buckled honeycomb structure. Light and dark blue balls represent the two vertically displaced sublattices. \textbf{b} The bandstructure of germanene with SOC (blue shaded area) and without SOC (dashed lines), depicting the topological gap and the two edge states with red and blue crossing the Fermi level. \textbf{c} Large-scale STM image of parallel germanene nanoribbons. Left side of the area has isolated nanoribbons, while the right side has an array of nanoribbons. The in-between regions are filled by nanowires and disorder domains (setpoints: 300 mV, 200 pA). \textbf{d} High-resolution image of several germanene nanoribbons recorded on a dense area (setpoints: -300 mV, 200 pA). \textbf{e} An atomic resolution STM image revealing the honeycomb lattice of the nanoribbon marked with red square in panel \textbf{c} (setpoints: 50 mV, 200 pA). \textbf{f} Atomic resolution image of a germanene nanoribbon showing the honeycomb lattice and the zigzag termination (setpoints: -300 mV, 200 pA). \textbf{g} line profile recorded across the honeycomb lattice of the nanoribbon shown with the black dashed line in the inset, revealing a buckling of 0.35 \AA. \textbf{h} Tentative structural model of germanene (red lattice) on Pt/Ge(110) substrate. 
}
\label{NR:Fig1}
\end{figure*}

\begin{figure*}[tb]
	\centering
	\includegraphics[width=0.95\textwidth]{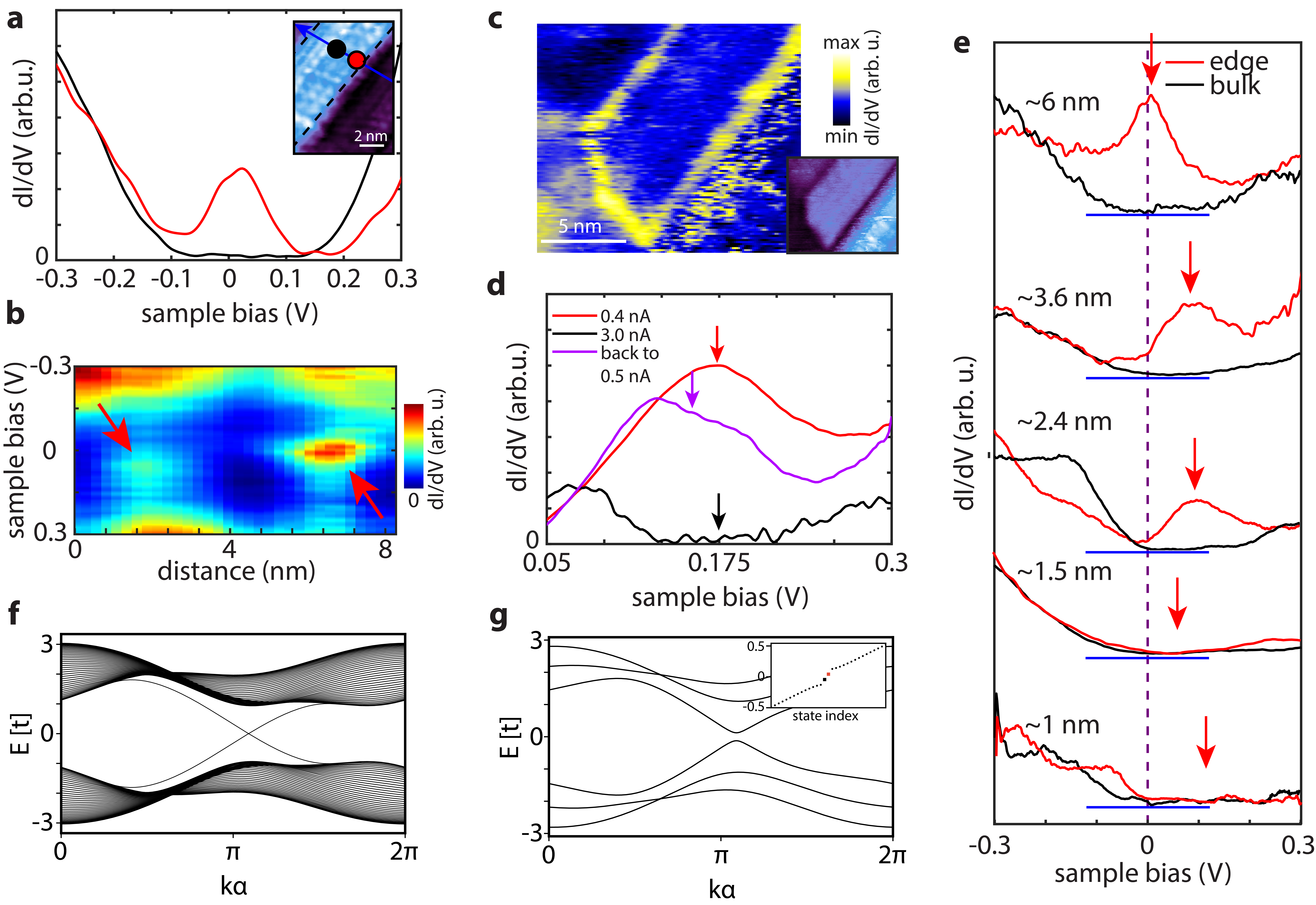}
	\caption{\textbf{Topologically protected edge states.} \textbf{a} $dI(V)/dV$ spectra at the edge (red) and interior (black) of the germanene nanoribbon, as shown in the inset (edges marked by black dashed lines). Measurement locations are indicated by red and black circles. Setpoints: -500 mV and 200 pA. \textbf{b} \textit{dI(V)/dV} line spectroscopy across the nanoribbon, recorded at the position of the blue arrow in \textbf{a}, showing clear edge states at both sides of the nanoribbon, highlighted by the red arrows. \textbf{c} \textit{dI/dV} map recorded at the energy location of the edge state (150 mV for this case) showing the edge state running continuously along the nanoribbon edge. Inset: the corresponding STM topography. \textbf{d} $dI(V)/dV$ spectra recorded at the edge of a wide nanoribbon for different current setpoints in chronological order: 0.4 nA (red), 3 nA (black), and 0.5 nA (magenta), showing the electric field-induced destruction of the topological state (the arrows indicate the edge state). \textbf{e} $dI(V)/dV$ spectra recorded at the edges (red) and interior (black) of nanoribbons of different widths (setpoints: -300 mV and 200 pA). Edge states vanish in ribbons narrower than 2.4 nm. Spectra are vertically offset for clarity, with blue horizontal lines indicating the zero levels. The red arrows indicate the edge state or the middle of the gap for narrow nanoribbons. \textbf{f} Calculated band structure for a 14-hexagon ribbon with crossing edge states. \textbf{g} Calculated band structure of a 2-hexagon ribbon, with gapped edge states and 0D end states marked by red/black squares in the inset.}
\label{NR:Fig2}
\end{figure*}

\begin{figure*}[tb]
	\centering
	\includegraphics[width=0.95\textwidth]{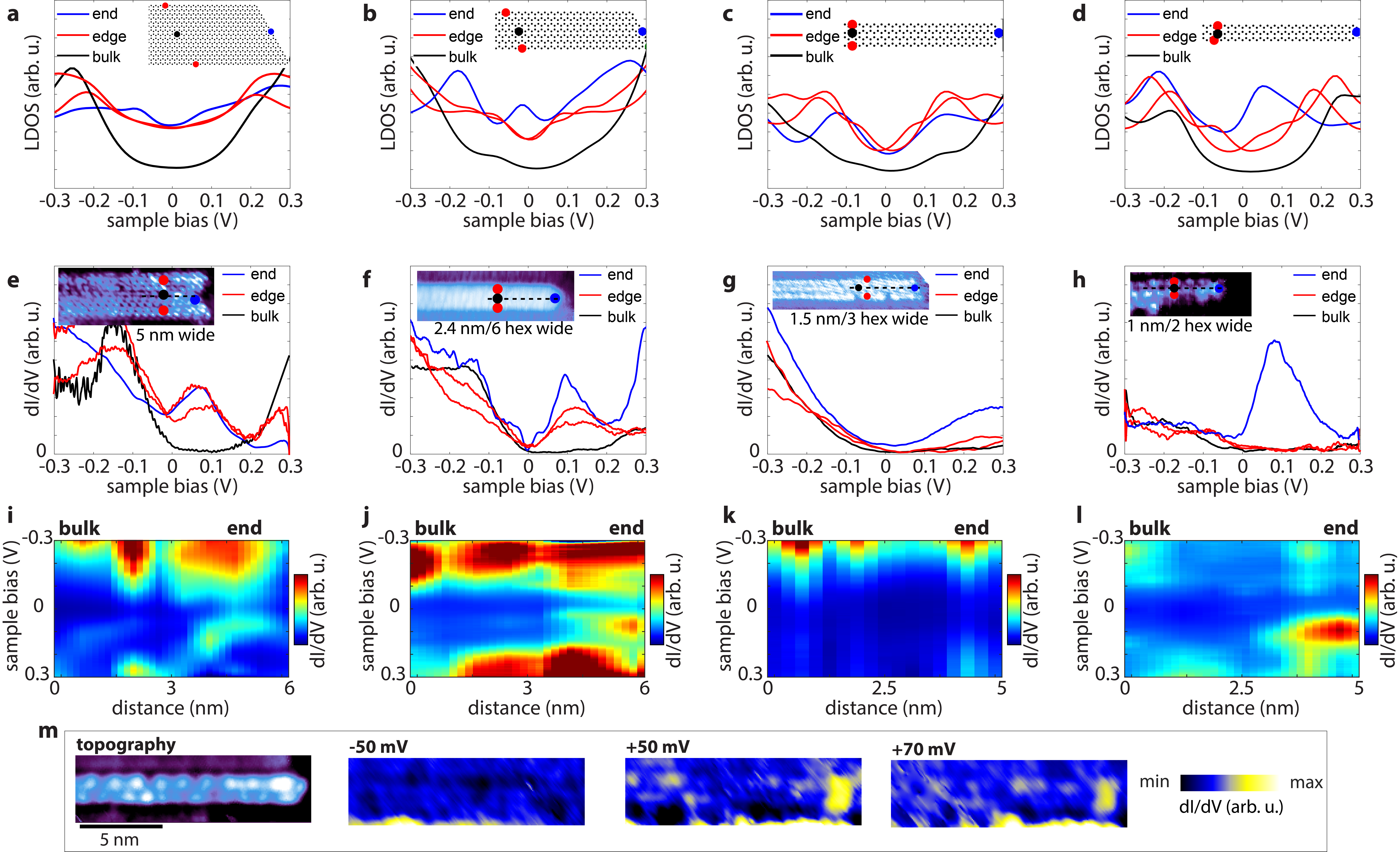}
	\caption{\textbf{Transition from a 2D TI to a 1D TI.} Panels \textbf{a}, \textbf{b}, \textbf{c}, and \textbf{d} show tight-binding calculations of the Local Density of States (LDOS) at the edges (red), end (blue), and bulk (black) for wide (14 hexagon units), intermediate (6 hexagon units), 3-hexagon and 2-hexagon wide nanoribbons, respectively. Insets illustrate nanoribbon structures. Panels \textbf{e}, \textbf{f}, \textbf{g}, and \textbf{h} present $dI(V)/dV$ spectra measured at the edges (red), end (blue), and bulk (black) for wide ($\sim$5 nm), intermediate ($\sim$2.4 nm, 6 hexagon units), a 3-hexagon ($\sim$ 1.5 nm), and 2-hexagon ($\sim$1.1 nm) wide nanoribbons, respectively. Insets show nanoribbon topographies and measurement locations. The data illustrate a reduction and disappearance of 1D topological edge states with decreasing width, and the emergence of 0D topological end states in agreement with theoretical calculations. Panels \textbf{i}, \textbf{j}, \textbf{k}, and \textbf{l} show the corresponding $dI(V)/dV$ line spectroscopy from the bulk to the end of the nanoribbons recorded at the locations marked with black dashed lines in the insets of panels \textbf{e}, \textbf{f}, \textbf{g}, and \textbf{h}. The \textit{dI(V)/dV} spectra were recorded with setpoints: -300 mV and 200 pA. \textbf{m} Topography and \textit{dI/dV} maps for different energies of a 2-hexagon ultrathin nanoribbon showing the end states and its localization at the end of the nanoribbon.}
\label{NR:Fig3}
\end{figure*}

\renewcommand{\thefigure}{S\arabic{figure}}



\clearpage
\newpage

\section*{{\Huge Supplementary Information for 'Realization of a one-dimensional topological insulator in ultrathin germanene nanoribbons'}}

\setcounter{figure}{0}
\newpage

\section{Experimental Methods}

\subsection{Scanning Tunneling Microscopy and Low Energy Electron Microscopy and Diffraction}
Scanning tunneling microscopy (STM) data were obtained with a UHV Omicron low-temperature STM operated at 77 K. The samples were transferred from the preparation chamber to the STM chamber without breaking the vacuum. The conductivity and sharpness of the tips were tested on either an Au or a Pt-coated HOPG sample. STM imaging was done with typical current setpoints in the range of 0.5-1 nA and the voltage biases were limited between -0.5 V and 0.5 V. The STS (\textit{dI(V)/dV}) data were obtained with the feedback loop open, using a lock-in amplifier with a modulation voltage of 10-20 mV AC at a frequency range of 1000-1200 Hz. The STM images and \textit{dI(V)/dV} spectra were processed and analyzed using MATLAB, Gwyddion, and SPIP. Low energy electron microscopy and diffraction (LEEM/LEED) experiments were performed using an ELMITEC LEEM-III instrument. The LEED patterns were recorded at 1150 K, using 30 eV electrons. 

\subsection{Nanoribbon Growth and Structure}

Germanene nanoribbons were synthesized using a two-step process: first, depositing approximately one monolayer of Pt on a Ge(110) substrate, and then annealing to 1150 K. Upon cooling, Ge segregates on the Pt monolayer, similar to the approach by Yuhara \textit{et al.} \cite{yuhara2021epitaxial}. For a monolayer of Pt on Ge(110), annealing to 1150 K causes Ge atoms to segregate on top of the Pt layer in order to maximize the Ge-Pt bonds \cite{gurlu2004initial, zhang2016structural2, zhang2018modified, watanabe2016pt}). The subsurface Pt layer conforms to the Ge(110) surface structure due to the good match between Pt(110) and Ge(110), with less than 2\% strain in both directions. This results in a surface structure similar to Ag on Ge(110) \cite{yuhara2021epitaxial}. After the annealing step, two types of structures emerge due to the segregation of Ge on top of the Pt-rich layer. These structures are contingent on the Pt coverage. An almost complete monolayer of Pt favors germanene nanoribbon growth, whereas lower local Pt coverages lead to Ge nanowire growth, with their density depending on the Pt coverage \cite{zhang2016structural2}, see Fig. S1. Both nanoribbons and nanowires run in the same direction ([-110]) due to crystal symmetry. Unlike the nanowires, the nanoribbons extend over hundreds of nanometers and have a more perfect and flat structure. The nanoribbons are characterized by a buckled honeycomb lattice. 

\begin{figure*}[h!tbp]
	\centering
	\includegraphics[width=0.8\textwidth]{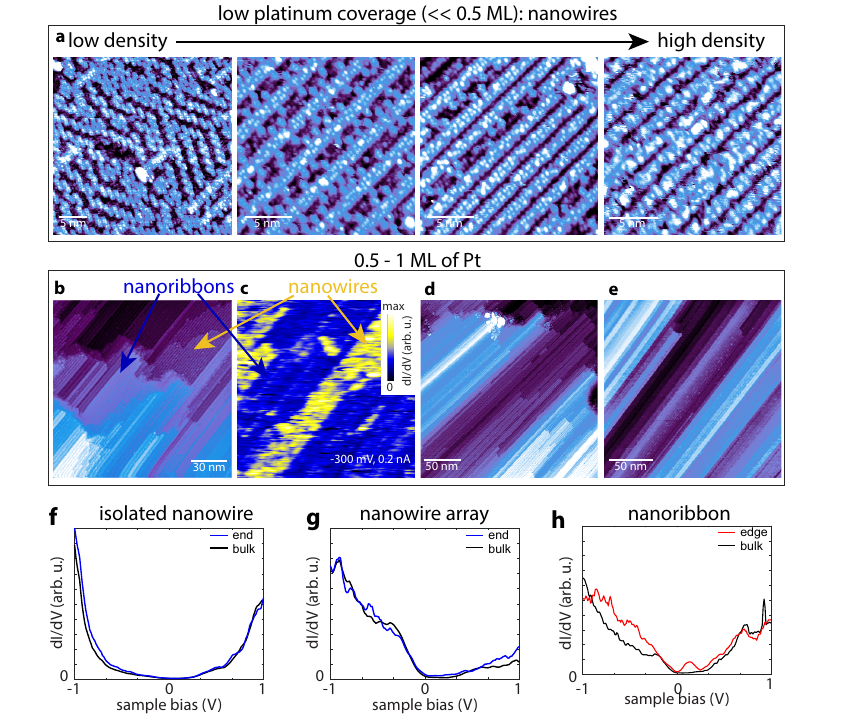}
	\caption{\textbf{a} Pt-induced nanowires formed after the deposition of small Pt amounts ($<$ 0.5 ML). The image shows the effect of Pt coverage on nanowire formation up to 0.5 ML. The nanowires are disordered without any regular long-range structure \textbf{b} For higher Pt coverages ($>$ 0.5 ML), nanoribbons, such as the ones shown in \textbf{b}, emerge. \textbf{c} The corresponding \textit{dI/dV} map of the topography in \textbf{b}, revealing that the nanowires are distinct compared to nanoribbons. \textbf{d} and \textbf{e} show different locations with large arrays of nanoribbons. \textbf{f} \textit{dI(V)/dV} spectrum of an individual nanowire, showing insulating behavior for both the bulk and the end. \textbf{g} \textit{dI(V)/dV} spectrum of a nanowire in a nanowire patch, exhibiting more metallic characteristics, though the spectra remain featureless at both the bulk and end. \textbf{h} \textit{dI(V)/dV} spectrum of a wide nanoribbon, highlighting the distinct edge and end state. 
 }
\label{NR:FigS1}
\end{figure*}

\subsubsection*{Coverage dependent nanoribbon growth}

The effect of Pt coverage on the surface structures formed on Ge(110) is shown in Fig. S1. For low Pt coverages (less than 0.5 monolayers, ML), Pt-induces the formation of Ge nanowires on the substrate. These nanowires are sparsely distributed initially but become denser as the Pt coverage increases, as observed when moving from left to right in Fig.~\ref{NR:FigS1}\textbf{a}. The nanowires have ill-defined structures without any periodicity. At higher Pt coverages, exceeding 0.5 ML, the surface composition shifts significantly, leading to the emergence of germanene nanoribbons instead of nanowires, first forming local patches and then covering almost the whole surface. These nanoribbons, shown in Fig. ~\ref{NR:FigS1}\textbf{b}, exhibit a distinct and periodic structure compared to the nanowire arrays seen at lower coverages. The differential conductance (\textit{dI/dV}) map in Fig. ~\ref{NR:FigS1}\textbf{c}, corresponding to the topography in Fig.~\ref{NR:FigS1}\textbf{b}, reveals that the nanowires and the nanoribbons can be distinguished from each other. This is because the nanoribbons have different electronic characteristics, allowing for a clear distinction between the two structures. Fig.~\ref{NR:FigS1}\textbf{d} and \textbf{e} present images from different samples, showcasing large arrays of long nanoribbons. These arrays demonstrate the consistency and reproducibility of nanoribbon formation at about a monolayer Pt coverages. The nanowires and nanoribbons exhibit clear spectroscopic differences. As shown in Fig. S1\textbf{f-h}, individual nanowires are insulating, while arrays of nanowires display more metallic characteristics, although their \textit{dI(V)/dV} spectra remain featureless. On the other hand, the nanoribbons have distinct properties that depend on their size as discussed in the main text. In Fig S1\textbf{h} the spectrum of a wide nanoribbon is shown for both bulk and edge, showing the edge state filling the bulk band gap.

\subsubsection*{Structure of germanene nanoribbons}

\begin{figure*}[h!tbp]
	\centering
	\includegraphics[width=0.7\textwidth]{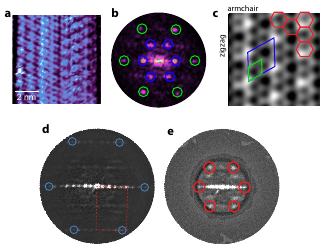}
	\caption{\textbf{a} High-resolution STM topography of a nanoribbon recorded in conditions such that both the ($1\times 1$) and ($2\times 2$) germanene lattices are visible at the same time. \textbf{b} FFT of \textbf{a} showing the ($1\times 1$, green circles) and ($2\times 2$, blue circles) germanene lattice structures. \textbf{c} FFT-filtered STM image of the nanoribbon in \textbf{a}, revealing the honeycomb lattice and zigzag termination along the [$\bar{1}$10] direction. \textbf{d} Cumulative $\mu$LEED pattern (electron energies between 1.5 and 21 eV) measured at room temperature in a region that contains nanoribbons. \textbf{e} $\mu$LEED pattern (electron energy 3 eV) measured at room temperature on a region that contains germanene nanoribbons. The blue and red circles refer to the ($1\times 1$) and ($2\times 2$) germanene lattices, respectively. The diffuse background of the inelastically scattered electrons is removed for every single image in the constituting series.}
\label{NR:FigS2}
\end{figure*}

The atomically resolved STM images in the main text reveal the low-buckled honeycomb lattice of the germanene nanoribbons with a lattice constant of $\sim$ 4.2 \AA. Depending on the sample bias (or the tip state), a larger hexagonal lattice is often also visible, see Fig.~\ref{NR:FigS2}\textbf{a-c}, with a lattice constant $2\times \alpha \sim$ 8.4 \AA.  Under specific tip conditions, both the honeycomb ($1\times 1$) and the ($2\times 2$) lattice are visible within a single STM image, see Fig.~\ref{NR:FigS2}\textbf{a} and the corresponding Fast Fourier Transformation (FFT) analysis in Fig.~\ref{NR:FigS2}\textbf{b}. This structure becomes very clear in the corresponding back-transformed image, Fig.~\ref{NR:FigS2}\textbf{c}. It is unclear what causes this larger periodicity at this stage, but it could be the effect of the germanene-substrate registry. Micro-spot Low-Energy Electron Diffraction ($\mu$LEED) imaging of a nanoribbon-containing area further confirms this, showcasing both the $2\times \alpha$ periodicity and the $1\times 1$ germanene lattice, Fig.~\ref{NR:FigS2}\textbf{d} and \textbf{e}. Unlike STM's local imaging, $\mu$LEEDs broader overview (1.4 $\mu$m spot size), incorporates regions that also contain nanowires with distinct 3x and 5x periodicity \cite{zhang2016structural2}, see Fig.~\ref{NR:FigS1}\textbf{a}, giving rise to the pronounced diffraction spots in Figs.~\ref{NR:FigS2}\textbf{d} and \textbf{e}. The nanowires cover most of the surface and align also with the [$\bar{1}$10] direction. Due to their smaller occupation area in this sample, nanoribbon-associated diffraction spots are thus less prominent. The cumulative $\mu$LEED pattern, covering energies from 1.5 to 21 eV, displays a broader specular beam along the [001] direction compared to the [$\bar{1}$10] direction, see Fig.~\ref{NR:FigS2}\textbf{d}, suggesting a dense step structure along the [$\bar{1}$10] direction, consistent with the presence of nanoribbons. $\mu$LEED analysis at 3 eV electron energies, Fig.~\ref{NR:FigS2}\textbf{e}, reveals the distinctive ($2\times 2$) hexagonal lattice (marked with red circles) of the nanoribbons. For completion, the ($1\times 1$) germanene lattice is visible in Fig.~\ref{NR:FigS2}\textbf{d} and marked with blue circles.

\pagebreak

\section{Tight-binding calculations}
\subsection{Topology in thin zigzag nanoribbons}
The tight-binding results for the zigzag nanoribbons are obtained by solving the Kane-Mele model with longer range hopping, 
\begin{equation}
    \begin{split}
        H = & \ t\sum_{\langle ij\rangle}c_i^\dagger c_j^{} + t_2\sum_{\langle\langle ij\rangle\rangle}c_i^\dagger c_j^{} + t_3\sum_{\langle\langle\langle ij\rangle\rangle\rangle}c_i^\dagger c_j^{}  + \frac{i\lambda_{SO}}{3\sqrt{3}}\sum_{\langle\langle ij\rangle\rangle} \nu_{ij}c_i^\dagger s_z c_j^{} \\
        &+ i\lambda_R\sum_{\langle ij\rangle}c_i^\dagger (\mathbf{s} \times \mathbf{d}_{ij})_z c_j^{} + M_S\sum_{i}\eta_i c_i^\dagger c_i^{},
    \end{split}
    \label{eq:HKM}
\end{equation}
on a nanoribbon geometry. Here, $c_i^{(\dagger)}$ is the electronic annihilation (creation) operator at site $i$, $t=-0.92$\ eV is the nearest-neighbor (NN) hopping parameter, and $t_2$ and $t_3$, respectively, the next-nearest neighbor (NNN) and next-next-nearest neighbor (NNNN) hopping parameters; $\lambda_{SO}$ and $\lambda_R$ are the intrinsic and Rashba SOC strength, and $M_S$ is a staggered on-site mass (Semenoff mass), which may account for the effect of a perpendicular electric field on buckled lattices, such as the one of germanene. Furthermore, $s_i$, $i\in \{x,y,z\}$, represent the Pauli spin matrices, $\nu_{ij}$ is the Haldane phase, which is +1 (-1) for (counter) clockwise hopping, $\mathbf{d}_{ij}$ is the vector connecting sites $i$ and $j$, and $\eta_i$ is a staggering parameter, which is +1 (-1) on the A (B) sublattice. Rashba SOC may arise because of broken inversion symmetry due to coupling to the substrate, for example. In the experiments, it is likely present, but we assume it to be small and neglect it. The staggered mass $M_S$ may be generated by the electric field resulting from the STM tip. It may also arise from a coupling between the ribbon and the substrate.
In the absence of Rashba SOC, spin is a good quantum number and the different spin blocks of the Hamiltonian may be treated independently. These are obtained by taking two time-reversal symmetric copies of the Haldane model at Haldane phase $\phi=\pi/2$. For this reason, it is sufficient to restrict our analysis to a single Haldane model, given by 
\begin{equation}
    H=t\sum_{\langle ij\rangle}c_i^\dagger c_j^{} + t_2\sum_{\langle\langle ij\rangle\rangle}c_i^\dagger c_j^{} + t_3\sum_{\langle\langle\langle ij\rangle\rangle\rangle}c_i^\dagger c_j^{} + \frac{i\lambda_{SO}}{3\sqrt{3}}\sum_{\langle\langle ij\rangle\rangle} \nu_{ij}c_i^\dagger c_j^{} + M_S\sum_{i}\eta_i c_i^\dagger c_i^{}. \label{eq:MHKM}
\end{equation}
We investigate the Hamiltonian in a ribbon geometry, i.e., we consider open boundary conditions in the $y$-direction and periodic boundary conditions in the $x$-direction.\\

Recently, Traverso \textit{et al.} did an in-depth analysis of the behavior of topological boundary states using the Haldane model for thin zigzag nanoribbons \cite{traverso2024emerging}. We follow their argument. The zigzag nanoribbons are quasi-one-dimensional, such that their 0D topological properties may be captured by the Zak phase $\varphi$ \cite{zak1989berry}. For multiband systems, it is given by
\begin{equation}
    \varphi = -\Im \log \det \prod_{j=0}^{N-1} S(k_j, k_{j+1}), \label{eq:Zak}
\end{equation} where $S(k_j, k_{j+1})$ is a matrix in momentum space with matrix elements given by the overlap of wavefunctions: 
$S_{mn}(k_j, k_{j+1}) \equiv \langle u_{m k_j} \mid u_{n k_{j+1}} \rangle$. Here, $k_j$ denotes the discretized momentum $k_j \equiv 2 \pi j / (a N)$, and the band indices $n,m$ are chosen to lie below the Fermi level. Moreover, the periodic gauge $\lvert u_{nk_N} \rangle = e^{-i \frac{2\pi}{a} \hat{x}} \lvert u_{nk_0} \rangle$ is enforced. The Zak phase implicitly depends on the system parameters $t$, $\lambda_{SO}$, $M_S$, and the width $N_y$ of the ribbon.

In the case of zigzag nanoribbons (ZZNR), the Zak phase on its own does not describe the phase of the system. This can be understood straightforwardly: In the limit of $M_S\to \infty$, the quantum spin Hall phase is absent. As a result, there are no 1D edge modes traversing the bulk gap. Without the 1D edge modes, there is no minigap, and consequently, the 0D topology is deemed to be trivial. Following this argument, we call a system with a certain Zak phase topological if its value is distinct from the value at $M_S \to \infty$: $\varphi_{M_S\to \infty}$. There is one final caveat: for systems of even width, the Zak phase takes different values in the limit $M_S\to \pm \infty$. This issue is resolved by comparing the Zak phase to $\varphi_{M_S\to \infty}$ if $M_s>0$, and to $\varphi_{M_S\to -\infty}$ if $M_S<0$. The topological invariant for 0D end modes may then be defined as
\begin{equation}
    \nu \equiv \frac{\varphi - \varphi_{M_S\to \pm\infty}}{\pi} \mod 2. \label{eq:topinvarSM}
\end{equation}

Finite $\nu$ indicates the presence of topological 0D end states located at the termination of the zigzag nanoribbons. The Zak phase is quantized as a result of the combined mirror and time-reversal symmetry, see section 2.2.

\begin{figure*}[h!tbp]
    \centering
    \includegraphics[width=\textwidth]{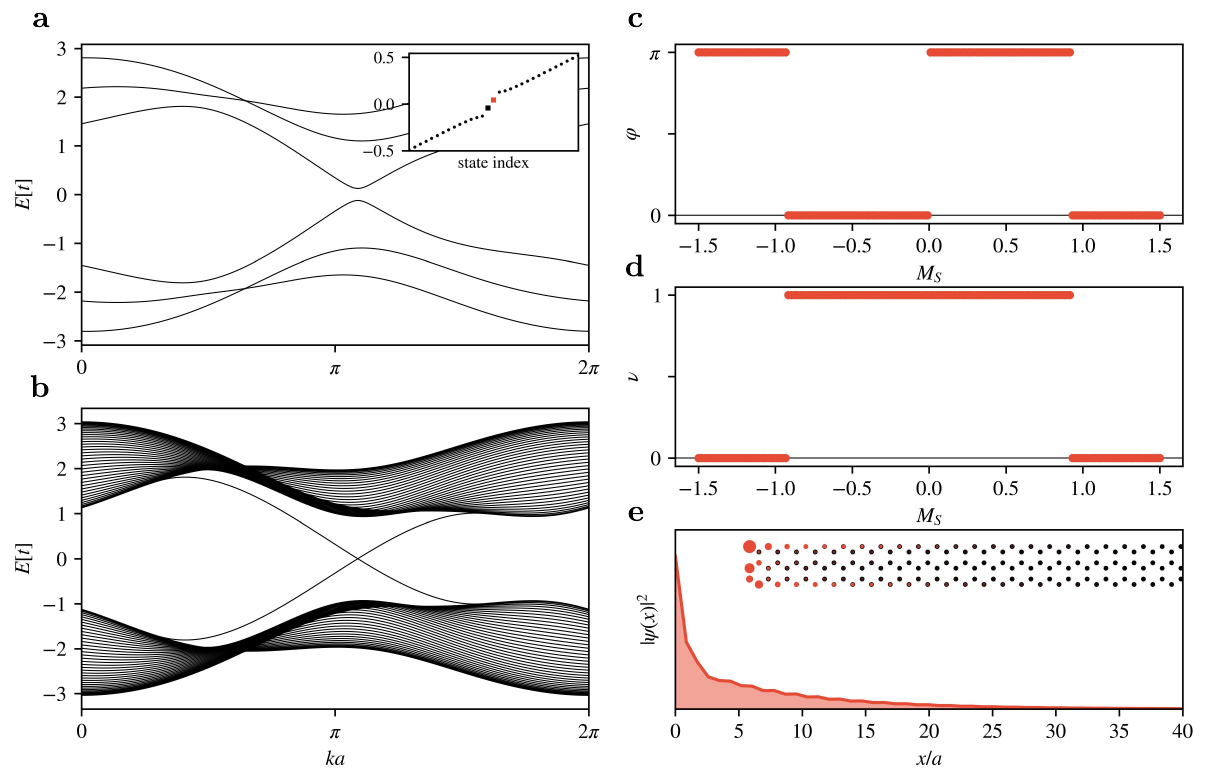}
    \caption{\textbf{a} Band structure of the thin Ge ZZNR with $t=1$, $\lambda_{SO}/3\sqrt{3}=0.3t$, $M_S=0.5t$. The inset shows a zoom in on the states inside the minigap for open boundary conditions. The two 0D end states are represented by a red and a black square. \textbf{b} Band structure of a wider Ge ZZNR for the same parameters. \textbf{c} Zak phase and \textbf{d} topological invariant of a thin Ge ZZNR. A value of 0 (1) indicates a trivial (topological) state. \textbf{e} Projected probability amplitude of the state indicated by a red square in the inset of \textbf{a}. The inset in \textbf{e} shows the unprojected probability density, where the radius of the red circles scales with $\lvert\psi\rvert^2$.}  
\label{fig:Methods}
\end{figure*}

\begin{figure}[H]
    \centering
    \includegraphics[width=0.7\textwidth]{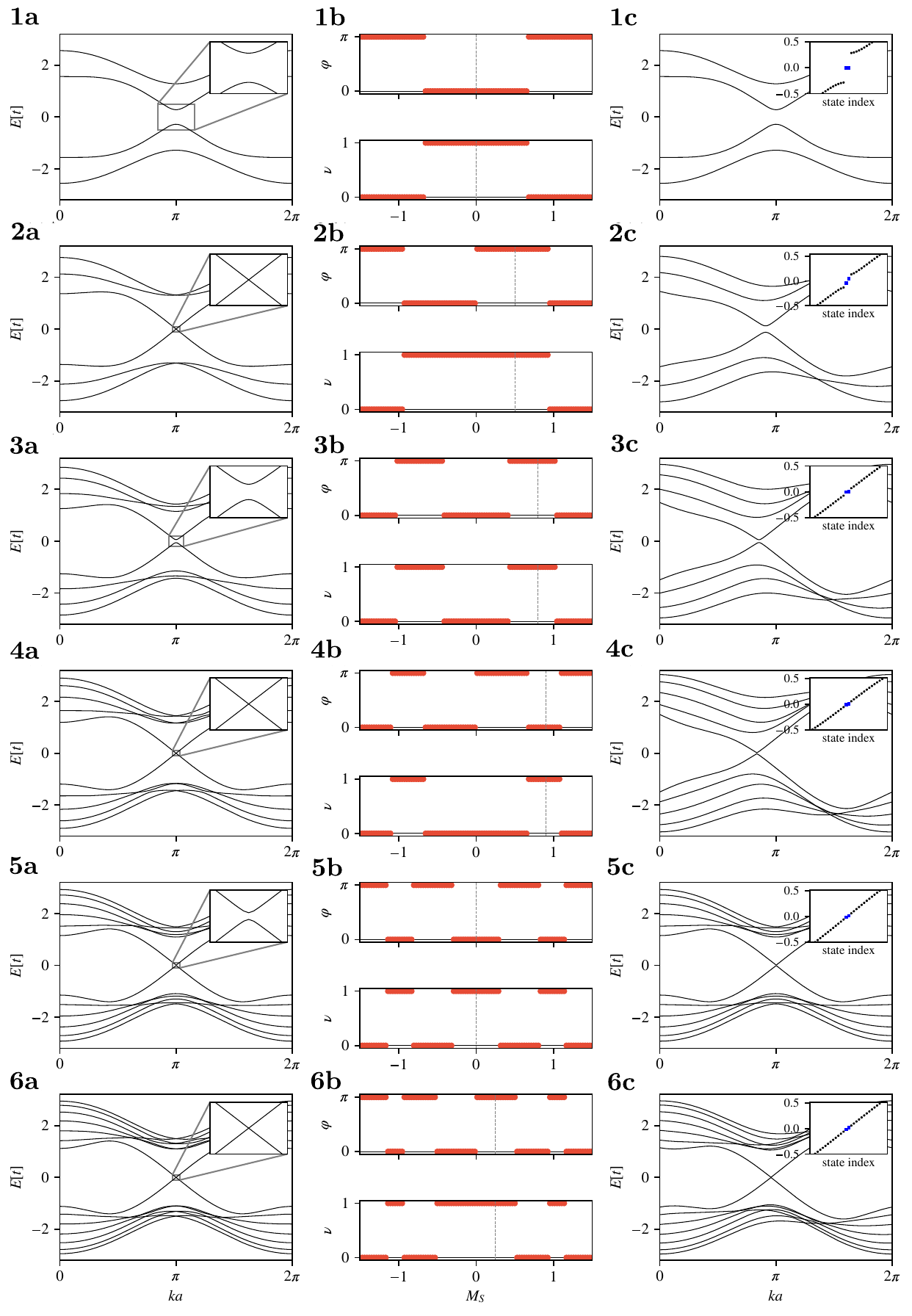}
    \caption{Width-dependent topology in thin Haldane ZZNRs. In all plots, $\lambda_{SO}/3\sqrt{3}=0.3t$. The rows indicate increasing ribbon width, starting at the top row with a 1-hexagon-unit wide nanoribbon, up to a 6-hexagon-unit wide nanoribbon for the bottom row. Left column: Nanoribbon band structure for $M_S=0t$. The inset shows a zoom in on the minigap. Middle column: Zak phase $\varphi$ and topological invariant $\nu$ as a function of $M_S$. Right column: Nanoribbon band structure for the value of $M_S$ indicated by dashed lines in the middle column. The values are chosen such that the ribbon is in the topological phase. The inset shows the open boundary condition spectrum around the minigap. 0D topological states are marked by blue squares.}
    \label{fig:evenodd}
\end{figure}

Fig.~\ref{fig:Methods}\textbf{a} elaborates on the 1D topology in the nanoribbons. For the sake of illustration, we have chosen exaggerated parameters here. In the case of open boundary conditions (OBC), in the topological phase, there are two in-gap modes. Fig.~\ref{fig:Methods}\textbf{b} shows an example of a wider ribbon, with gap traversing edge modes. Figs.~\ref{fig:Methods}\textbf{c-d} show, respectively, the Zak phase $\varphi$, calculated using Eq.~\ref{eq:Zak}, and the topological invariant $\nu$ defined in Eq.~\ref{eq:topinvarSM}, for a 2-hexagon wide ribbon. Fig.~\ref{fig:Methods}\textbf{e} shows how the in-gap states are exponentially localized at the ends of the ribbon. To furhter investigate the relation between ribbon width and topology, we turn to Fig.~\ref{fig:evenodd}. The different rows indicate different ribbon widths, with the width increasing from 1 hexagon-unit (top row) to 6 hexagon units (bottom row). The left column shows the band structure for $\lambda_{SO}/3\sqrt{3}=0.3t$ and $M_S=0$. In the case of vanishing mass, the ribbons of odd width (odd number of hexagon units) are gapped, and the gap size decreases upon increasing the ribbon width. On the contrary, ribbons of even width are gapless.

The middle column shows the Zak phase \cite{zak1989berry} $\varphi$ and the associated topological invariant $\nu$ at $\lambda_{SO}/3\sqrt{3}=0.3t$ as a function of $M_S$. Consider, for example, row 2 of Fig.~\ref{fig:evenodd}. For very large negative $M_S$, the Zak phase is $\pi$, which corresponds to the trivial phase. Upon increasing $M_S$, the Zak phase drops to zero, which is different from the limit at large negative $M_S$, indicating a topological phase. When $M_S$ crosses zero, the Zak phase jumps to $\pi$, which is different from the Zak phase for very large positive $M_S$. Thus, this is a topological phase. It should be noted here that precisely at $M_S=0$ there is no gap, and the Zak phase is ill-defined. For large (positive) $M_S$, the Zak phase drops back to zero, indicating a trivial phase.

The third column of Fig.~\ref{fig:evenodd} shows the band structure of the ribbon at the value of $M_S$ indicated by the dashed lines in the plots of $\nu$ (middle column in Fig.~\ref{fig:evenodd}). The values are chosen such that the nanoribbon is in the topological phase. For this parameter value, there is a minigap, albeit small. In the case of open boundary conditions, there are two topological end modes within the minigap, indicated by the blue squares in the inset. For increasing width, the minigap shrinks and the topological modes will merge with the 1D edge modes. 

\begin{figure*}[h!tbp]
    \centering
    \includegraphics[width=\textwidth]{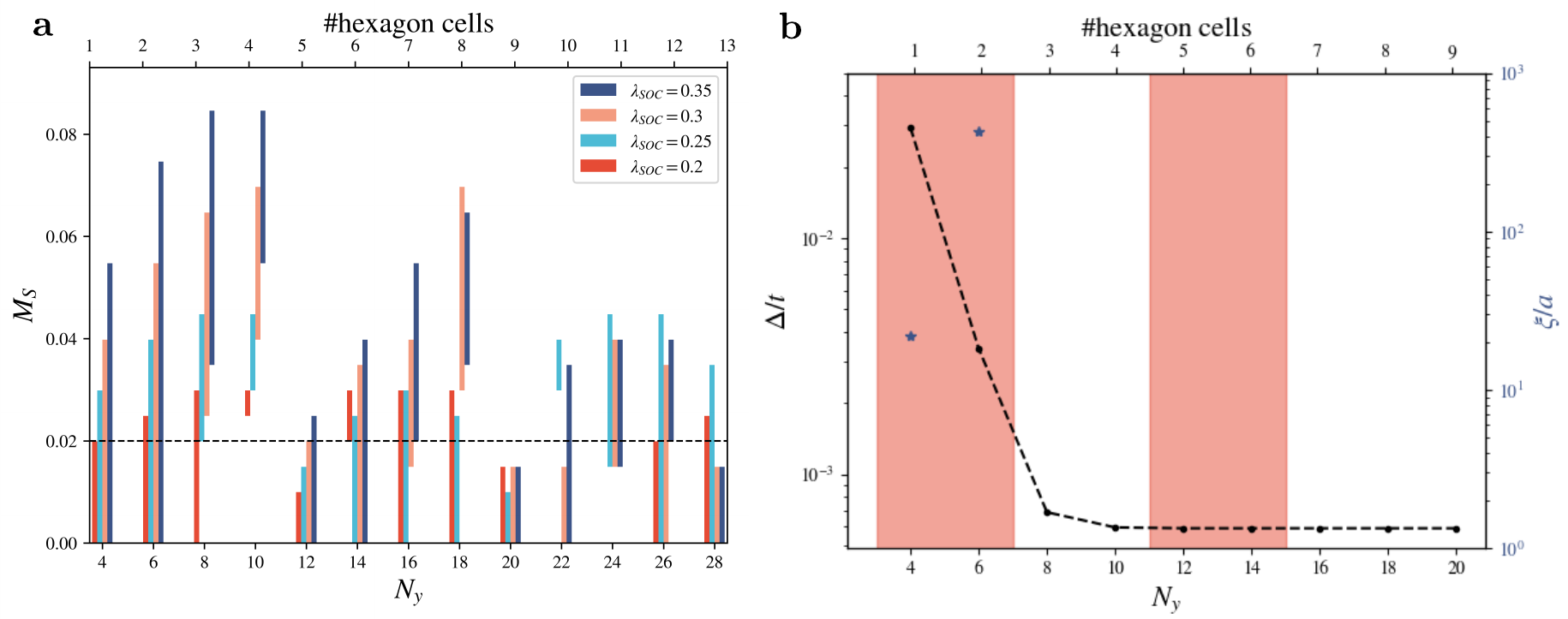}
    \caption{\textbf{a} Phase diagram for the Ge zigzag nanoribbon (ZZNR) as a function of the staggered mass $M_S$ and width $N_y$. Colored regions indicate non-trivial ($\nu=1$) topology, manifesting as 0D topological end modes. Each color indicates a different spin-orbit coupling strength. \textbf{b} Minigap size $\Delta$ (black) and 0D end mode localization length $\xi$ (blue stars) along the dashed line indicated in \textbf{a} for $t_2=t_3=0$ and $\lambda_{SO}=0.35t$. Red regions indicate that the gap is topological and hosts 0D end modes.}
    \label{fig:combinedphase}
\end{figure*}

A phase diagram of the 0D topology in zigzag nanoribbons for different values of $\lambda_{SO}$ is shown in Fig.~\ref{fig:combinedphase}\textbf{a}. Notice that the system exhibits re-entrant topological behavior as a function of its width. For example, consider the dark blue region $(\lambda_{SO}=0.35t$) at a fixed staggered mass $M_S=0.02t$ (horizontal dashed line in Fig. S1\textbf{a}). The system is topological for $N_y=4$ and $6$. It then becomes trivial until a width of $N_y=12$ is reached, where it is topological once again, before it becomes trivial for $N_y>14$. Similar results have also been obtained experimentally. A nanoribbon of 3 hexagonal units ($N_y = 8$) without end states is shown in Fig. 3\textbf{g}, which is in agreement with the tight-binding results presented in Fig. 3\textbf{c}. 

For thin enough nanoribbons, the topological 1D edge modes fail to fully close the bulk gap, leading to a minigap. The 0D end modes sit within this minigap. Fig.~\ref{fig:combinedphase}\textbf{b}  shows the minigap size as a function of ribbon width at fixed SOC strength $\lambda_{SO}=0.35t$ and fixed staggered mass $M_S=0.02t$ (dashed line in Fig. S1\textbf{a}). The gap size decreases upon increasing the ribbon width, while the end mode localization length increases (blue stars). Notice that the vertical axes are logarithmic. For the wider topological nanoribbons ($N_y=12$), the gap size becomes so small that the end modes can no longer be resolved for the chosen parameters. In this case, the end modes could be regarded as having merged into the edge.

\subsection{Quantization of the Zak phase}
The topology in (Haldane) zigzag nanoribbons is protected by a combined mirror $\mathcal{M}_x$, and time-reversal symmetry $\mathcal{T}$. This can be understood as follows: time-reversal flips the direction of the complex next-nearest neighbour hopping by changing the Haldane phase. The application of mirror symmetry then sends $x$ to $-x$, which flips the direction of this hopping again and sends the Hamiltonian back to itself. Below, we show how combined $\mathcal{M}_x\mathcal{T}$ symmetry quantizes the Zak phase.\\

The multi-band Zak phase,
\begin{equation}
    \varphi = -\Im \log \det \prod_{j=0}^{N-1} S(k_j, k_{j+1}), \label{eq:ZakTrav}
\end{equation}
with overlap matrix elements $S_{mn}(k_j,k_{j+1}) = \braket{u_{mk_j}}{u_{nk_{j+1}}}$, is equal to the sum of the arguments of the eigenvalues $\exp{i\theta_\alpha}$ of the Wilson loop
\begin{equation}
    \mathcal{W}_{2\pi \xleftarrow{}0} =  \prod_{j=0}^{N-1} S(k_j, k_{j+1}), \label{eq:Wilson}
\end{equation}
i.e. $\varphi = \sum_\alpha \theta_\alpha \mod 2\pi$.\\

\noindent The action of the different relevant symmetries on the Wilson loop spectrum is:
\begin{itemize}
    \item \noindent In the case of time-reversal symmetry, we have
\begin{align*}
    \mathcal{T} \mathcal{W}_{2\pi \xleftarrow{} 0 } \mathcal{T}^{-1} = \mathcal{W}_{0 \xleftarrow{} 2\pi } ^*  = \mathcal{W}_{2\pi \xleftarrow{} 0 } ^T.
\end{align*}
This puts no constraints on the Wilson loop, as an operator and its transpose share the same eigenvalues.\\
\item \noindent In the case of mirror-symmetry, we have
\begin{align*}
    \mathcal{M}_x \mathcal{W}_{2\pi \xleftarrow{} 0 } \mathcal{M}_x^{-1} = \mathcal{W}_{0 \xleftarrow{} 2\pi }  = \mathcal{W}_{2\pi \xleftarrow{} 0 } ^\dagger.
\end{align*}
This imposes the condition $\exp{i\theta_\alpha} = \exp{-i\theta_\alpha}$ on the Wilson loop eigenvalues, i.e. $\theta_\alpha = 0,\pi$.\\

\item \noindent Finally, under the combined action of mirror and time-reversal symmetry, we have
\begin{align*}
    \left(\mathcal{M}_x \mathcal{T} \right) \mathcal{W}_{2\pi \xleftarrow{} 0 } \left(\mathcal{T}^{-1}\mathcal{M}_x^{-1}\right) = \mathcal{M}_x \mathcal{W}_{0 \xleftarrow{} 2\pi } ^*\mathcal{M}_x^{-1}  = \mathcal{W}_{2\pi \xleftarrow{} 0 } ^*.
\end{align*}
This again imposes the condition $\exp{i\theta_\alpha} = \exp{-i\theta_\alpha}$ on the Wilson loop eigenvalues, i.e. $\theta_\alpha = 0,\pi$.
\end{itemize}
\noindent Both in the presence of $\mathcal{M}_x$ and $\mathcal{M}_x \mathcal{T}$, the Wilson eigenvalues, and consequently the Zak phase can only take values $0$ and $\pi$. Armchair nanoribbons do not have this symmetry and therefore do not exhibit end states. The end states in zigzag nanoribbons are protected by $\mathcal{M}_x \mathcal{T}$ symmetry.

\subsection{Persistence of 0D topological phase}
To better understand the robustness of the 0D end states in the thin nanoribbon, Fig.~3 of the main text, we extensively analyzed the LDOS of thin (2- and 3-hexagon units), intermediate width (6-hexagon units), and wide ribbons (14-hexagon units). By diagonalizing the Hamiltonian (Eq.~\ref{eq:HKM}), we obtain its spectrum $\{E_n\}$ and eigenstates $\{ \lvert \psi_n(\mathbf{r}) \rangle \}$. This allows us to numerically determine the LDOS,
\begin{equation}
    \text{LDOS}(E,\mathbf{r}) = \frac{1}{\pi} \sum_n \lvert\psi_n(\mathbf{r})\rvert^2 \frac{b}{(E-E_n)^2+b^2},
\end{equation}
where $b$ accounts for broadening effects. In our calculations, we take $b = 0.05$\ eV.

\begin{figure}[H]
    \centering
    \includegraphics[width=\textwidth]{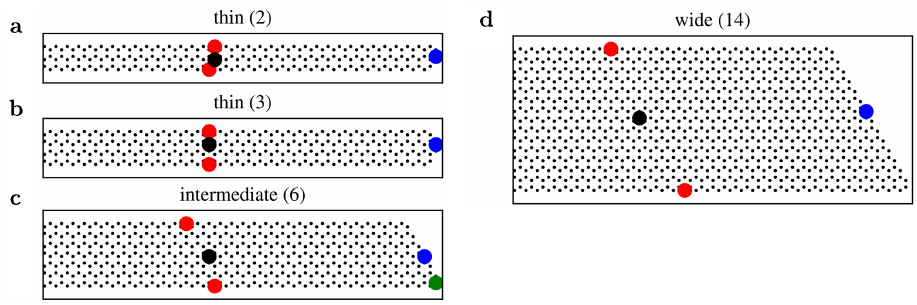}
    \caption{Geometry of the nanoribbons considered in the main text. Colored dots denote the positions at which the LDOS curves in Figs.~(\ref{fig:S4}-\ref{fig:S5}) are taken.}
    \label{fig:ribbons}
\end{figure}

Fig.~\ref{fig:ribbons} shows the geometry of ribbons of different width. The colored points denote the positions at which the LDOS curves shown in Figs.~\ref{fig:S4}-\ref{fig:S5} are taken. In the case of the intermediate ribbon (Fig.~\ref{fig:ribbons}\textbf{c}), we indicate two positions at the end, blue and green. This is done because the increase of LDOS at the end of the 6-hexagon nanoribbon, Figs.~3\textbf{b} and 3\textbf{f} of the main text, is the net result of two effects. It can partly be attributed to the ribbon termination. When the ribbon ‘turns’ by 120 degrees, there is a single armchair unit. This turning point is indicated by the green dot. The blue dot indicates the mid-point of the short boundary side. It is slightly off-centered because we chose it to be one of the zigzag ‘teeth’, which does not always sit exactly at the center of the short boundary side. Tight-binding calculations predict a slight increase in LDOS at this position. For comparison, in Figs.~\ref{fig:S4} and Figs.~\ref{fig:S5}, for a 6-hexagon ribbon, we show the end LDOS both at the turning point (green) and in the middle of the end side (blue). The green curve has a higher intensity, but both exhibit a clear LDOS peak. In addition, there is a 0D topological end mode, which yields an increase in LDOS around E = 0 for both the blue and green curves.\\ 

\begin{figure}[H]
    \centering
    \includegraphics[width=\textwidth]{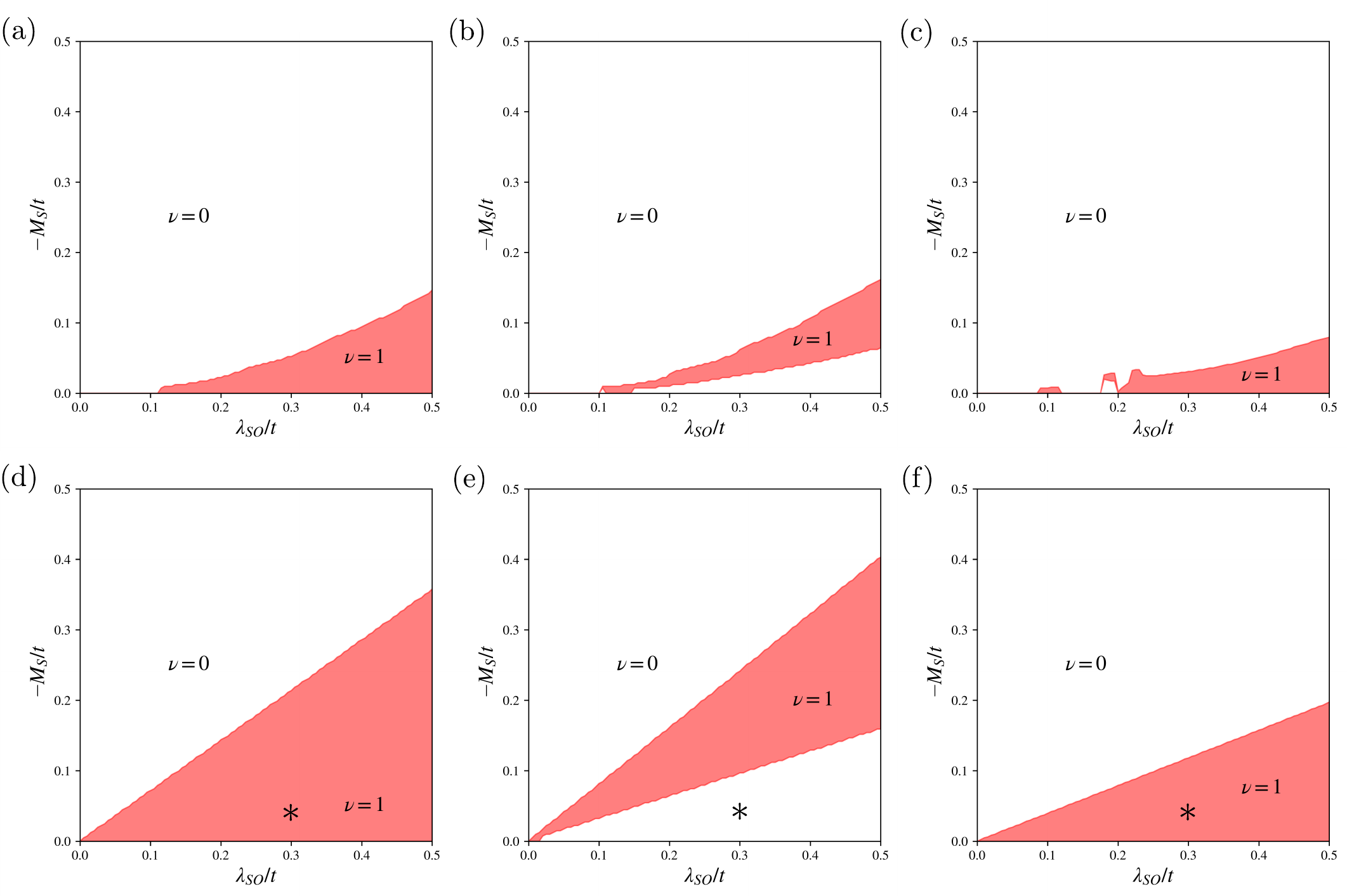}
    \caption{Phase diagrams of the nanoribbons shown in Figs.~\ref{fig:ribbons}\textbf{a}-\textbf{c}. \textbf{a}, \textbf{d} The phase diagrams are shown for the 2-hexagon thin nanoribbon; \textbf{b}, \textbf{e} a 3-hexagon thin nanoribbon; \textbf{c}, \textbf{f} a 6-hexagon (intermediate-width) nanoribbon. The top row has only nearest-neighbor hopping, while the bottom row has $t_3=0.3t$.  $\nu=0$ ($\nu=1$) represents a trivial (topological) phase. In the topological phase, exponentially localized 0D end modes are present at the termination of the nanoribbon. Stars indicate the parameters chosen in the main text.}
    \label{fig:combined}
\end{figure}

\begin{figure}[H]
    \centering
    \includegraphics[width=0.75\textwidth]{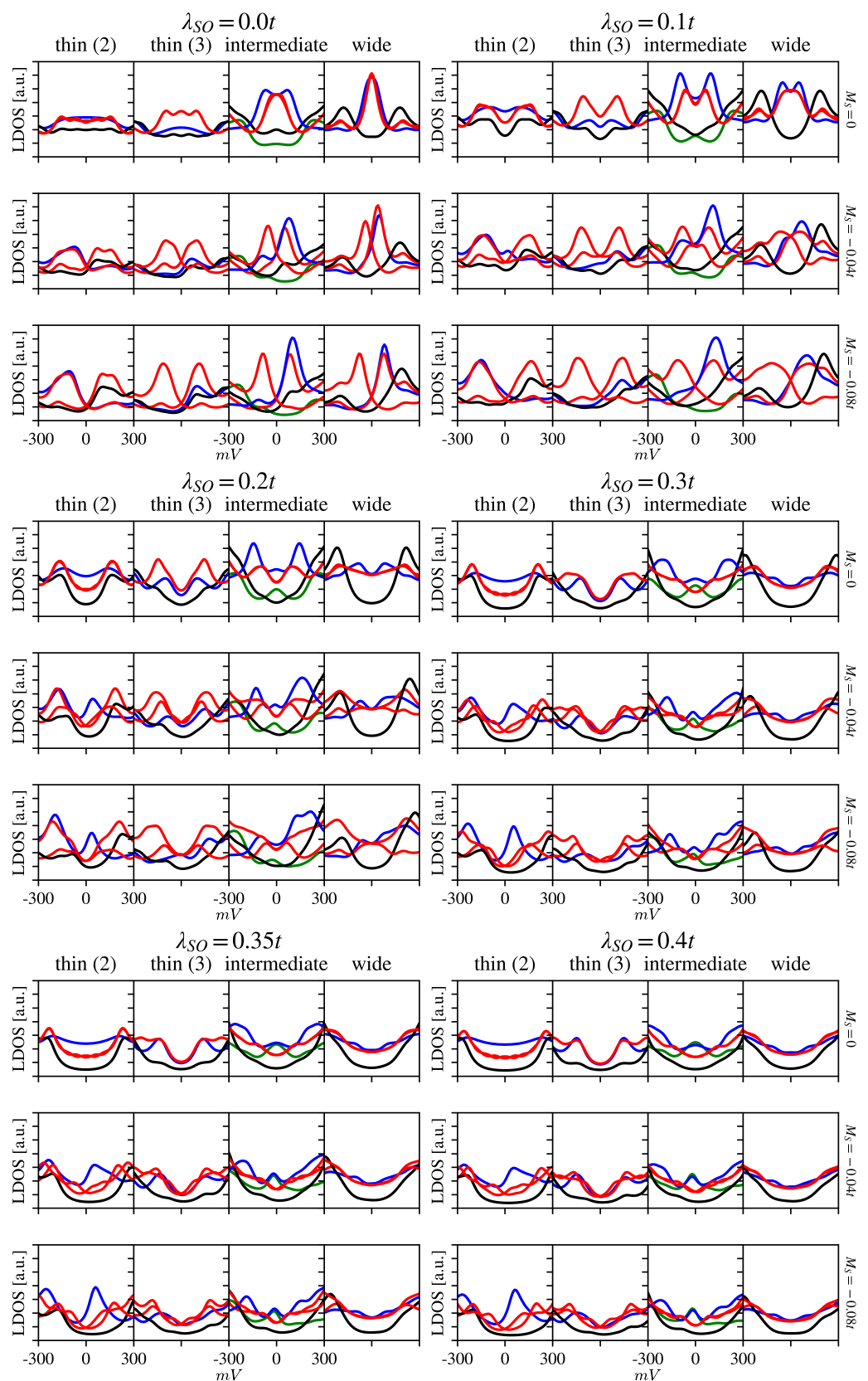}
    \caption{LDOS of the thin (2- and 3-hexagon units), intermediate width, and wide nanoribbons for different spin orbit strengths, $t=-0.92 eV$, $t_2=0, t_3=0.3t$.  $M_S$ ranges from $0t$ to $-0.08t$, and the different colored lines represent the position at which the LDOS curve is taken, see Fig.~\ref{fig:ribbons}.}
    \label{fig:S4}
\end{figure}

\begin{figure}[H]
    \centering
    \includegraphics[width=0.75\textwidth]{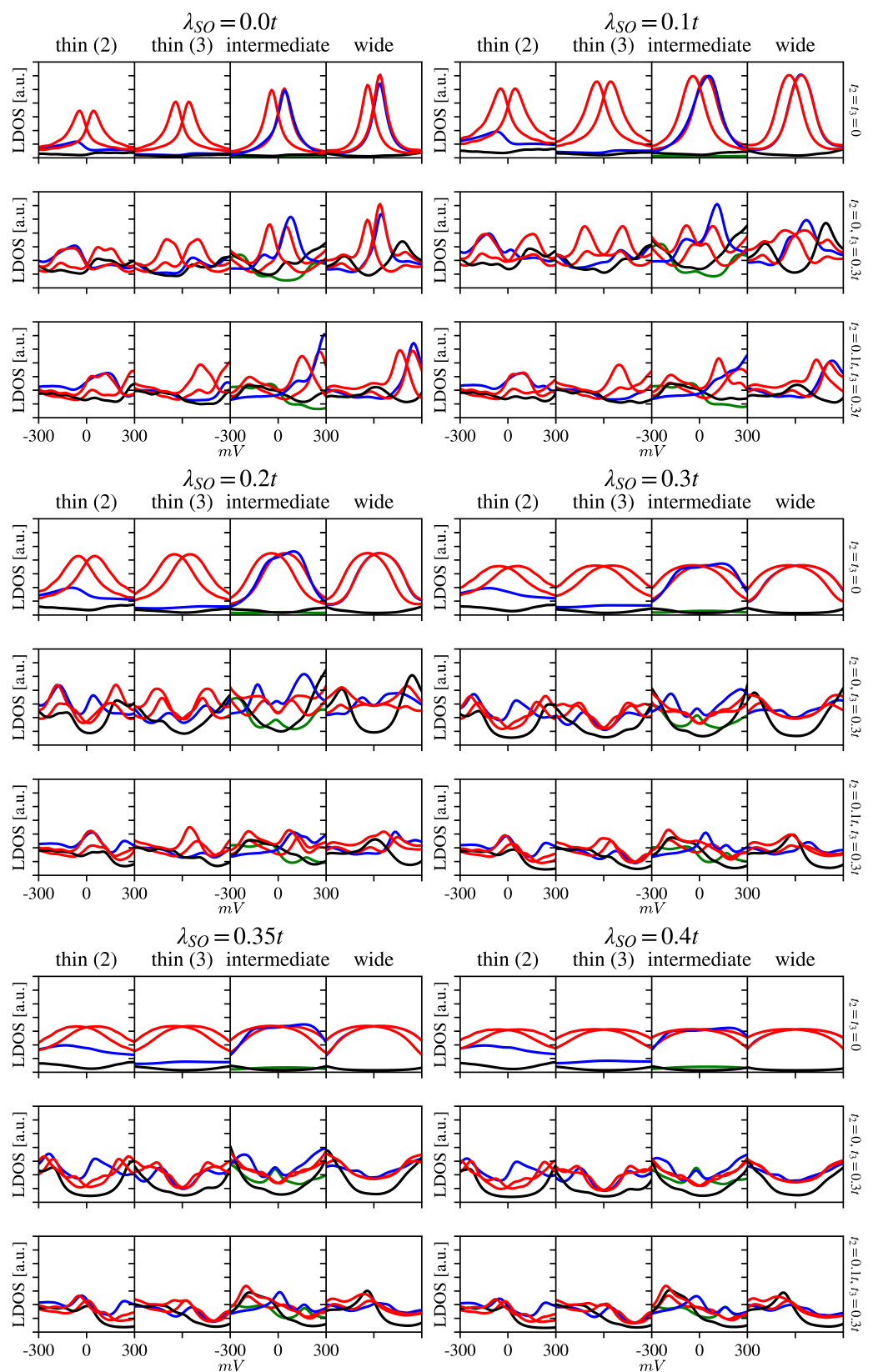}
    \caption{LDOS of the thin (2- and 3-hexagon units), intermediate width, and wide nanoribbons for different SOC strength and different longer range hopping amplitude. $M_S=-0.04t$ and the different colored lines represent the position at which the LDOS curve is taken, see Fig.~\ref{fig:ribbons}.}
    \label{fig:S5}
\end{figure}

Fig.~\ref{fig:combined} shows phase diagrams for the thin and intermediate width nanoribbons. The top row corresponds to the case where there is no long-range hopping, while the bottom row corresponds to $t_3=0.3t$. Figs.~\ref{fig:combined}\textbf{a} and \ref{fig:combined}\textbf{d} show the phase diagrams of the thin (2) ribbon depicted in Fig.~\ref{fig:ribbons}\textbf{a}, Figs.\ref{fig:combined}\textbf{b} and \ref{fig:combined}\textbf{e} represent the thin (3) ribbon, and Figs.~\ref{fig:combined}\textbf{c} and \ref{fig:combined}\textbf{f} represent the intermediate ribbon.

By analyzing the entire regime of parameters, as in Fig.~\ref{fig:S4}, we can observe several features. First, in the absence of SOC, the end modes, characterized by a blue peak at zero bias, never arise. Indeed, even in thin (2) ribbons in the presence of a Semenoff mass, the blue and red curves stay on top of each other, and no peak emerges. Surprisingly, the presence of a strong SOC seems to be a necessary but not sufficient condition to the emergence of the zero bias peak because even in the presence of strong SOC, there is no blue peak when the Semenoff mass is zero, although the blue and red curves no longer retrace each other. It is therefore essential to have a finite Semenoff mass together with the intrinsic SOC to observe the (blue) peak in the thin ribbons. A third prominent feature is that there is a strong difference in behaviour for thin ribbons with even or odd number of hexagon-units in width. For $\lambda_{SO}=0.3t$ and $M_S=-0.08t$, e.g., there is a peak for the thin (2) ribbon but not for the thin (3) ribbon. From $\lambda_{SO}$ of order of $0.2t$ and above, the blue peak is always visible around zero energy at finite $M_S$ (the peak gets slightly shifted away from zero due to the interplay among $M_S$,  SOC, and $t_3$). This effect becomes better defined as the strength of the Semenoff mass increases.

Moreover, for $\lambda_{SO}=0$ and $M_S=0$, there are clear (symmetric) red peaks and blue peaks in Fig.~\ref{fig:S4} (see, for example, the top row of the figure for all widths). These are a consequence of the zigzag termination of honeycomb lattices, which can host non-topological edge modes. For finite $M_S$, these peaks move away from zero and still correspond to the non-dispersive non-topological zigzag edge states. 

A direct comparison between Figs.~\ref{fig:combined} and \ref{fig:S4} yields a good agreement. Consider for example $\lambda_{SO}=0.3t$: a peak is visible in the LDOS at the end of the thinnest ribbon (blue line, left column of Fig.~\ref{fig:S4}) for $M_S=-0.04t$ and $M_S=-0.08t$. Figure~\ref{fig:combined}\textbf{d} confirms that in this regime $\nu=1$. The peak does not appear when $M_S=0$, as in this case the minigap would not be opened. 

We now shift our focus to the influence of long-range hopping. References \cite{reich2002tight, chegel2020tunable} show that next-nearest-neighbor hopping $t_2$ and next-next-nearest-neighbor hopping $t_3$ can be as large as $30\%$ of the nearest-neighbor hopping, $t$, in graphene \cite{reich2002tight} and is sizable in germanene \cite{chegel2020tunable}. For this reason, we also include them in our model. 
Let us focus first on the effect of $t_3$. When inspecting Fig.~\ref{fig:combined}, one observes that the introduction of a finite $t_3$ enlarges the range of parameters for which the system is topological (compare, e.g., Figs.~\ref{fig:combined}\textbf{a} and \ref{fig:combined}\textbf{d}). Figure~\ref{fig:S5} further shows the influence of longer-range hopping for different spin-orbit strengths at fixed $M_S=-0.04t$. $t_3$ increases the size of the end state peak, while $t_2$ seems to shift it, as can be seen for $\lambda_{SO}=0.35t$. This can be understood by considering that $t_2$ connects sites in the same sublattice, breaking sublattice symmetry. Going from the first to second row yields a more prominent blue peak in the first column, i.e. a more prominent end mode. By increasing $t_2$, this peak shifts. 
In the main text, we compared the experimental results to the theoretical ones obtained for the parameters $\lambda_{SO}=0.3t$ and $M_S=-0.04t$. From Figs.~\ref{fig:combined}\textbf{d} and \ref{fig:combined}\textbf{e}, we conclude that for these parameters only the thin (2) ribbon should be topological. Figure~\ref{fig:S4} confirms this result, since only the thinnest ribbon shows a distinct peaked LDOS at the boundary (blue curve). Therefore, we conclude that it is likely that the thinnest experimental ribbons are only 2-unit cells wide.  

\end{document}